\documentclass[12pt]{article}
\usepackage{graphicx}
\begin{document}

\newcommand{\Dh}{\hbox{\bf D}}
\newcommand{\Eh}{\hbox{\bf E}}
\newcommand{\Fh}{\hbox{\bf F}}
\newcommand{\fh}{\hbox{\bf f}}
\newcommand{\gh}{\hbox{\bf g}}
\newcommand{\Gh}{\hbox{\bf G}}
\newcommand{\Hh}{\hbox{\bf H}}
\newcommand{\Lh}{\hbox{\bf L}}
\newcommand{\mh}{\hbox{\bf m}}
\newcommand{\Oh}{\hbox{\bf O}}
\newcommand{\Ph}{\hbox{\bf P}}
\newcommand{\Sh}{\hbox{\bf S}}
\newcommand{\uno}{\hbox{\bf 1}}
\newcommand{\dpslash}{\partial\kern-7pt /}
\newcommand{\Dhsl}{\Dh\kern-8pt /}
\newcommand{\Phsl}{\Ph\kern-8.6pt /}
\newcommand{\psl}{p \kern-4.5pt /}
\newcommand{\biprime}{\prime\prime}


\title{\vskip1cm
Free fermionic propagators on a lattice
\vskip.5cm}
\author{Rafael G. Campos$^*$ and Eduardo S. Tututi$^\dagger$\\ 
Escuela de Ciencias F\'{\i}sico--Matem\'aticas, \\
Universidad Michoacana \\
58060 Morelia, Michoac\'an, M\'exico\\
$^*$E-mail: rcampos@zeus.umich.mx\\
$^\dagger$E-mail:  tututi@zeus.umich.mx\\
Fax number: 52 443 316 6969
}
\date{}
\maketitle
{
}\\
\vspace*{4truecm}
\begin{center} Abstract \end{center}
A method used recently to obtain a discrete formalism for classical 
fields with nonlocal actions preserving chiral symmetry and uniqueness 
of fermion fields yields a discrete version of Huygens' principle with 
free discrete propagators that recover their continuum forms in certain limit. 
\vfill
\newpage

\section{Introduction}
Lattice theory has been used since many years ago as one of the 
non-perturbative approaches to study physical effects that could 
occur in QCD or QED \cite{Wil74}. Nevertheless, the discrete formalism yielded by 
the Lattice theory does not preserve important properties such as 
chirality and uniqueness which are present in the standard field 
theory. Nielsen and Ninomiya \cite{Nie81} have
shown that for a local and translationally invariant hermitian 
discretization of the fields it is not possible 
to have simultaneously chiral symmetry and uniqueness. Many attempts 
have been made to circumvent this restriction. The use 
of nonlocal operators can be found among others (see for example
\cite{Tin01}--\cite{Sla01}) in spite that the locality of the Dirac 
operator is in general a desired property. In this context, a new 
approach to the problem of discretization of fields is 
presented in \cite{CamTut1}. This approach is based on a non--equispaced 
lattice, performed through the zeros of Hermite polynomials, where both
discrete derivatives and Fourier transform have support \cite{Cam00a, Cam92}. 
The technique yields a projection of the quantum algebras
on a finite linear space yielding matrix representations for the partial
derivatives that produce discrete hermitian actions with nonlocal kinetic 
terms. The fermion doubling and the chiral symmetry breaking are absent
in such a formulation. In this work we obtain a discrete version of 
Huygens' principle for free spinor fields in $1+1$ D. We show that 
whenever the number of the nodes in the non--equispaced lattice tends 
to infinity, the propagator we found approaches to their continuum form.
We also show that the discrete propagator is indeed a discrete Green 
function by taking the inverse of the Dirac operator.\\
This work is organized as follows. In section \ref{secdis} we review
the essentials of our discretization method and introduce a discrete
version of the sign and step functions. In section \ref{sechuy} we 
obtain the discrete Huygens' principle on the non--equispaced lattice. 
Section \ref{propag} is devoted to study the discrete version of the
free Dirac propagator in $1+1$ D. Finally in section \ref{conclus} we
give our main conclusions.
\\


\section{Discrete technique} \label{secdis}

\subsection{Review of the method}
In this subsection we only present the main results of our discretization 
scheme; proofs and further applications can be found in \cite{CamTut1, 
Cam00a, Cam92, Cam00b, Cam00c, Cam97}.\\
Let us consider the non--equispaced four dimensional lattice constructed
with the set of nodal points $x^{\mu}_j$ ($\mu$ denotes the Lorentz index
and  $j=1,2,\ldots, N_{\mu}$, with $N_{\mu}$ being the number of nodes along
the direction $\mu$) performed by the zeros of the Hermite polynomial 
$H_{N_{\mu}}(\xi)$. We denote $N=N_0N_1N_2N_3$ as the number of lattice 
points. From this set of points we construct the four $N\times N$ matrices: 
\begin{equation}{
{\Dh}_0=D_0\otimes 1_{N_3}\otimes 1_{N_2}\otimes 1_{N_1},\qquad
{\Dh}_3=1_{N_0}\otimes D_3\otimes 1_{N_2}\otimes 1_{N_1}, }\atop{
{\Dh}_2=1_{N_0}\otimes 1_{N_3}\otimes D_2\otimes 1_{N_1}, \qquad
{\Dh}_1=1_{N_0}\otimes 1_{N_3}\otimes 1_{N_2}\otimes D_1,} \label{uno}
\end{equation}
where $1_{N_\mu}$ is the identity matrix of dimension $N_\mu$ and
$D_\mu$ is the skew-symmetric matrix 
\[
(D_\mu)_{jk}=\cases{0,&{$i=j$},\cr\noalign{\vskip .5truecm}
\displaystyle {1\over{x^\mu_j-x^\mu_k}}, &{$i\not=j$}.\cr} 
\]
Let us define the matrix ${\Sh}=S_0\otimes S_3\otimes S_2\otimes S_1$
where 
$(S_\mu)_{jk}=\delta_{jk}\exp[-(x^{\mu}_j)^2/2]H'_{N_{\mu}}(x^{\mu}_j)$.
Then, ${\Sh}{\Dh}_\mu{\Sh}^{-1}$ is a projection of the partial derivative $\partial_\mu$ in 
the subspace of functions $U$ generated by products of the form 
$ u_n(\xi)=\exp(-\xi^2/2)H_n(\xi)$, $n=0,1,\ldots,N_{\mu}-1$,
with $\xi=x^\mu$ and $\mu=0,1,2,3$. This means that such matrices are 
{\it exact} representations of the partial derivatives 
for functions in $U$. Therefore, whenever $N_\mu\to\infty$ we will 
get convergent approximations to the partial derivatives of a function 
$\psi(x)$ spanned by the basis $\{u_n(\xi)\}_0^\infty$. 
Thus, the discrete version of a dynamical differential variable
operating on such functions is essentially the matrix obtained under 
the replacement  $\partial_\mu\to{\Sh}{\Dh}_\mu{\Sh}^{-1}$.
The error arising from 
this procedure can be estimated in special cases \cite{Cam97} and can 
be related to the complement of $\psi(x)$ with respect to $U$ \cite{CamTut1}. 
Thus, if $\psi(x)\in U$ and $\Psi$ denotes the $N\times 1$ vector of components
\begin{equation}
\Psi_q=\psi(x_q)\equiv\psi(x^1_j,x^2_k,x^3_l,x^0_m), \label{fiq}
\end{equation}
ordered according to $q=j + (k-1)N_1 + (l-1)N_1N_2 + (m-1)N_1N_2N_3 $
where first $j$ runs over $j=1,\ldots,N_1$, then $k$ over
$k=1,\ldots,N_2$, then $l$ over 
$l=1,\ldots,N_3$, and finally we take $m=1,\ldots,N_0$.  
In this form we have that the vector $\Psi_{,\mu}$ constructed 
with the values of $\psi_{,\mu}=\partial_\mu\psi$ at the site 
$x_q=(x^0_m,x^1_l,x^2_k,x^1_j)$, is given by
\begin{equation}
\Psi_{,\mu}={\Sh}{\Dh}_\mu{\Sh}^{-1}\Psi. \label{cinco}
\end{equation} 
The similarity transformation given by ${\Sh}$ changes $\Psi$ 
into itself except for an alternating change of sign along each 
direction when $N_\mu\to\infty$  \cite{CamTut1}, 
therefore we may use ${\Dh}_\mu$ instead ${\Sh}{\Dh}_\mu{\Sh}^{-1}$ as 
a discrete representation of $\partial_\mu$. \\
Let $g(x^\mu)$ be a given function and $g$ and $g'$ the vectors of 
components $g(x^\mu_j)$ and $\partial g(x^\mu_j)/\partial x^\mu$, 
respectively. Let us denote by $G=\hbox{diag}(g)$ the diagonal matrix whose 
nonzero elements are the components of $g$. Then, we have that \cite{CamTut1}
\[
(S_\mu D_\mu S_\mu^{-1})G=G(S_\mu D_\mu S_\mu ^{-1})+G'+R_\mu, 
\]
where $G'=\hbox{diag}(g')$ and $R$ is the residual matrix which projects 
an arbitrary vector on the orthogonal subspace generated by the complement 
basis  $\{u_n(\xi)\}_{N_\mu}^\infty$. 
This equation, applied to the vector constructed with the values of the 
function $h(x^\mu)$ at the nodes, is the finite representation of the 
familiar formula $\partial(gh)/\partial x^\mu=g\partial h/\partial x^\mu+
h\partial g/\partial x^\mu$. Since the complement basis becomes empty 
as $N_\mu$ goes to infinite, we have that 
\begin{equation}
D_\mu G=GD_\mu +G', \qquad N_\mu\to\infty, 
\label{derprod}
\end{equation}
where $G'$ can be substituted by the diagonal matrix $\hbox{diag}(D_\mu g)$.
\\
In summary, ${\Dh}_\mu$ becomes a discrete representation of $\partial_\mu$ yielding
convergent results for a wide class of functions. However, this operator is not
a local one from the field theory point of view \cite{Tin01,Pil99}, i.e., there exist 
nonzero elements of ${\Dh}_\mu$ for which
\begin{equation}
\vert ({\Dh}_\mu)_{qq'}\vert > \exp(-\Vert x_q-x_{q'} \Vert_\infty) 
\label{dmexp}
\end{equation}
since $1/\vert z\vert > \exp(-\vert z \vert)$.  Here $\Vert\cdot\Vert_\infty$ is
the $\max$ vector norm. In Sec. \ref{sechuy} we will see a consequence
of this property.
\\
Now, let us define the symmetric function
\begin{equation}
F(\xi,\eta)=\sum_{l=0}^{N-1} (i)^l \varphi_l(\xi) \varphi_l(\eta),
\label{fxp1}
\end{equation}
where
$\varphi_l(\xi)=\bigl( 2^{N-1-l}(N-1)!/(N l!)\bigr)^{1/2} H_l(\xi)/ H_{N-1}(\xi).$
Since 
\begin{equation}
i{\Dh}_0 {\Fh}={\Fh}{\Ph}^0, \qquad -i{\Dh}_j {\Fh}={\Fh}{\Ph}^j, 
\label{dereig}
\end{equation}
the commuting matrices (\ref{uno}) can be diagonalized simultaneously
by the unitary and symmetric matrix
\begin{equation}
{\Fh}=F_0^\dagger\otimes F_3\otimes F_2\otimes F_1 \label{seis},
\end{equation}
where $(F_\mu)_{jk}=F(x^\mu_j,p^\mu_k)$ is a discrete Fourier transform 
for one variable \cite{Cam92}. 
Here, $p^\mu_j$ is also a zero of $H_{N_\mu}(\xi)$ and it represents an 
eigenvalue of the discretized momentum. 
The elements of the four--dimensional discrete Fourier transform (\ref{seis}) 
satisfy the important asymptotic formula 
\begin{equation}
\lim_{N\to\infty} {\Fh}_{q'q}=C_N\,e^{-ip_q\cdot x_{q'}}, \label{fasi}
\end{equation}
where 
$p_q\cdot x_{q'}=p^0_mx^0_{m'}-p^3_lx^3_{l'}-p^2_kx^2_{k'}-p^1_jx^1_{j'}$
and $C_N$ is the product of the constants 
$C_{N_\mu}=2^{{N_\mu}-3/2}(\Gamma[({N_\mu}+1)/2])^2/{N_\mu}!$ 
except for an alternating change of sign \cite{Cam00a, CamTut1}. By using 
an asymptotic expression for the Hermite zeros we find that $C_{N_\mu}$  
becomes proportional to the standard measure of a Riemann integral in
each variable (the difference between two consecutive lattice points):
\begin{equation}
C_{N_\mu}={1\over{\sqrt{2\pi}}}\Delta x^\mu=
{1\over{\sqrt{2\pi}}}\Delta p^\mu,
\label{mes}
\end{equation}
where $\Delta p^\mu=\Delta x^\mu=\pi/\sqrt{2N_\mu}$.
The equations (\ref{fasi}) and (\ref{mes}) are the backbone of 
the quadrature formula for the Fourier transform \cite{Cam92}.\\
Additional and useful properties of $F_\mu$, are
\begin{equation}
F_\mu^\dagger=(-1)^{N_\mu+1}F_\mu U, \qquad F_\mu U=UF_\mu,
\label{propif}
\end{equation}
where $U$ is the $N_\mu\times N_\mu$ matrix whose entries are given by
$U_{jk}=\delta_{j,N_\mu-k+1}$. Denoting by $[\cdot]_k$ the $k$th column 
of a matrix, the first equality means that 
\begin{equation}
[F_\mu]_k \to(-1)^{N_\mu+1}[F_\mu^\dagger]_k,  \label{fos}
\end{equation}
under $p^\mu_k\to -p^\mu_k$, $\mu=0,1,2,3$.

\subsection{Discrete distributions}
In section \ref{propag} a representation of the step function $\theta(t-t')$ 
on the lattice will be needed to construct the discrete propagators. 
In order to do that, let us give a discrete form of the sign function 
$\epsilon(t-t')$. Is it well known that 
\[
\epsilon(t-t')=-{i\over\pi}\int^\infty_{-\infty}
{{e^{i(t-t')E}dE}\over{E}}=-{i\over\pi}\int^\infty_{-\infty}
e^{itE}{1\over{E}}e^{-it'E}dE.
\]
According to (\ref{fasi}) and (\ref{mes}) the integral with the integrand 
$e^{itE}$ becomes the matrix $\sqrt{2\pi}F_0$ 
applied to the remaining part of the integrand. Thus, the discrete form 
of $\epsilon(t-t')$ is the skew-symmetric matrix 
\[
\Xi=-{i\over\pi}
(\sqrt{2\pi}F_0)(P^0)^{-1}({{\sqrt{2\pi}}\over{\Delta E}}F_0^\dagger)=
-{{2i}\over{\Delta E}}F_0(P^0)^{-1}F_0^\dagger,
\]
where $P^0$ is the diagonal matrix whose nonzero elements are $E_k$. Of 
course, $\Xi$ has to be a nonsingular matrix, restricting us to consider 
$N_0$ even. The fact that the entries of $\Xi$ are real 
is guaranteed by (\ref{propif}), and (\ref{fasi})--(\ref{mes}) 
yields a finite asymptotic value for any fixed element\footnote{By a fixed 
element of a $N\times N$ matrix $A$, we mean the entry $A_{N/2+j,N/2+k}$, 
with $j$ and $k$ fixed. Here $N$ is even and $j,k=\pm 1,\pm 2,\ldots, \pm N/2$.} 
of $\Xi$. Since 
$D_0F^*_0=-iF^*_0P^0$, the discrete derivative of $\Xi$ is two times the identity 
matrix divided by the measure, i.e.,
\begin{equation}
D_0\Xi=2({1\over{\Delta E}})1_{N_0\times N_0} \label{derdel}.
\end{equation}
The measure goes to zero as $N_0\to\infty$, therefore, this result is the 
discrete version of $d\epsilon(t-t')/dt=2\delta(t-t')$.
>From (\ref{derdel}) 
we see that $-iF_0(P^0)^{-1}F_0^\dagger$ is the inverse of $D_0$ and
that $D_0\Xi-\Xi D_0=0$.
\\
On the other hand, the step function accepts the Fourier representation 
\begin{equation}
\theta(t-t')=-{i\over{2\pi}}\int^\infty_{-\infty}
{{e^{i(t-t')E}dE}\over{E-i\epsilon}}=-{i\over{2\pi}}\int^\infty_{-\infty} 
{{e^{i(t-t')E}dE}\over{E}}+{\epsilon\over{2\pi}}\int^\infty_{-\infty}
{{e^{i(t-t')E}dE}\over{E^2}}.
\label{cotet}
\end{equation}
This equation is just the simple relation $\theta(t)=[\epsilon(t)+1]/2$. Since the 
first positive zero of $H_{N_0}(E)$ (the first positive value of 
$E_k$, i.e, $E_{{{N_0}\over{2}}+1}$) goes to zero as $N_0\to\infty$, 
we may give a discrete representation of $\theta(t-t')$ through 
(\ref{cotet}). The identification of $\epsilon$ with 
$E_{{{N_0}\over{2}}+1}$ gives the matrix 
\begin{equation}
\Theta={1\over2}(\,\Xi+\Sigma), \label{tetd}
\end{equation}
as the representation of $\theta(t-t')$ on the lattice, where
we have defined 
\[
\Sigma={{2E_{N_0/2+1}}\over\Delta E}F_0(P^0)^{-2}F_0^\dagger.
\]
A fixed element of this symmetric matrix 
becomes asymptotically equal to $1$ divided by the measure, i.e., 
\begin{equation}
(\Sigma)_{jk}={{1}\over{\Delta E}},\qquad N_0\to\infty.
\label{sigjk}
\end{equation}
Note that the discretization of (\ref{cotet}):
\[
-{{i}\over{\Delta E}}F_0(P^0-iE_{{{N_0}\over{2}}+1}1_{N_0\times N_0})^{-1}F_0^\dagger
\]
and (\ref{tetd}) are equal up to the first order in $E_{{{N_0}\over{2}}+1}$.
The application of $D_0$ to $\Sigma$ yields $-E_{{{N_0}\over{2}}+1}\Xi$. Since a 
fixed element of $\Xi$ is bounded as $N_0\to\infty$ and 
$E_{{{N_0}\over{2}}+1}\to 0$, we get the expected property 
\begin{equation}
D_0\Theta=\delta,
\label{deritet}
\end{equation}
where $\delta$ is the identity matrix divided by the measure, i.e.,
$\delta=1_{N_0\times N_0}/\Delta E$. In order to normalize the right--hand 
side of (\ref{deritet}) is convenient to our purpose define 
\begin{equation}
\tilde{\Theta}=
-iF_0(P^0-iE_{{{N_0}\over{2}}+1}1_{N_0\times N_0})^{-1}F_0^\dagger
={1\over2}(\,\tilde{\Xi}+\tilde{\Sigma})
\label{thetap}
\end{equation}
as the discrete form of $\theta(t-t')$, where
$\tilde{\Xi}=(\Delta E)\Xi$ and $\tilde{\Sigma}=(\Delta E)\Sigma$. 
In this way we have that 
$D_0\tilde{\Theta}=1_{N_0\times N_0}$. Another important relation 
between these matrices follows from (\ref{derprod}). If we choose the 
vector $g$ as the $j$th column of $\tilde{\Theta}$, i.e. 
$[\tilde{\Theta}]_j$, Eq. (\ref{derprod}) becomes
\begin{equation}
D_0 \hbox{diag}([\tilde{\Theta}]_j)=
\hbox{diag}([\tilde{\Theta}]_j)D_0+[1_{N_0\times N_0}]_j,\qquad N_0\to\infty, 
\label{derprodtet}
\end{equation}
\\

\section{Huygens' Principle} \label{sechuy}
The above formalism is applied in Ref. \cite{CamTut1} to find discrete 
spinor fields and some of their properties. In particular, the discrete 
version of the free Dirac equation is found to be
\begin{equation}
i\gamma^\mu\otimes {\Dh}_\mu\Psi=m {\bf 1}_{4N}\Psi, \label{eqdld}
\end{equation}
where $\gamma^\mu$ are Dirac matrices, ${\bf 1}_{4N}$ is the identity 
matrix of dimension $4N$ and $\Psi$ is the discretized field whose four 
spinorial components $\Psi_a$ have spatial--temporal indexes 
$(\Psi_a)_q=\Psi_a(x_q)$ ordered according to (\ref{fiq}). 
We remark that the Dirac operator ${\Dh}=i\gamma^\mu\otimes {\Dh}_\mu$  
is not a local one. To show this, let us 
take the matrix norm induced by the $\max$ vector norm \cite{Hor91}
\[
\Vert A_{N\times N} \Vert_\infty=N \max _{1\le j,k\le N}\vert a_{jk}\vert .
\]
Thus, according to (\ref{dmexp}), we have that there exist nonzero elements of
${\Dh}$ for which
\begin{equation}
\Vert {\Dh}(x_q,x_{q'}) \Vert_\infty > \max_{0\le\mu\le 3} \vert ({\Dh}_\mu)_{q,{q'}} \vert > 
 \exp(-\Vert x_q-x_{q'} \Vert_\infty). \label{normd}
\end{equation}
Accordingly, the no--go theorem due to Nielsen and Ninomiya does not apply 
in this case. Since the the condition for chiral symmetry is satisfied trivially 
by ${\Dh}$ \cite{CamTut1}, this operator yields chiral fermions with no doublers.
\\ 
The square of a mass eigenvalue of (\ref{eqdld}), say $m_r$, is given by
the discrete form of the energy--momentum relation 
\begin{equation}
m_r^2=(p^0_m)^2-(p^1_j)^2-(p^2_k)^2-(p^3_l)^2, 
\label{masq}
\end{equation}
where $p^\mu_i$ is a nonzero component of the diagonal matrix 
${\Ph}^\mu$ representing the discretized momentum $p^\mu$. Such values
are zeros of Hermite polynomials and therefore, $m_r$ is degenerate. 
The indexes in (\ref{masq}) are ordered according to 
\[
r=j + (k-1)N_1 + (l-1)N_1N_2 + (m-1)N_1N_2N_3 + (i-1)N_1N_2N_3N_0
\]
where the slowest index is $i=1,2,3,4$. The degeneracy of $m_r$ does 
not depend on $N_\mu$ and the discrete mass--shell 
consists only in few points even when $N_\mu\to\infty$. However, 
since the Hermite zeros become dense on the axes, $m_r^2$ approaches 
to any real number as $N_\mu\to\infty$ and its degeneracy is given 
by the number of points sufficiently close to the mass--shell defined
by $m_r$. This behavior is illustrated in Fig. 1.
\vskip1cm
\hbox to \textwidth{\hfill\scalebox{0.8}{\includegraphics{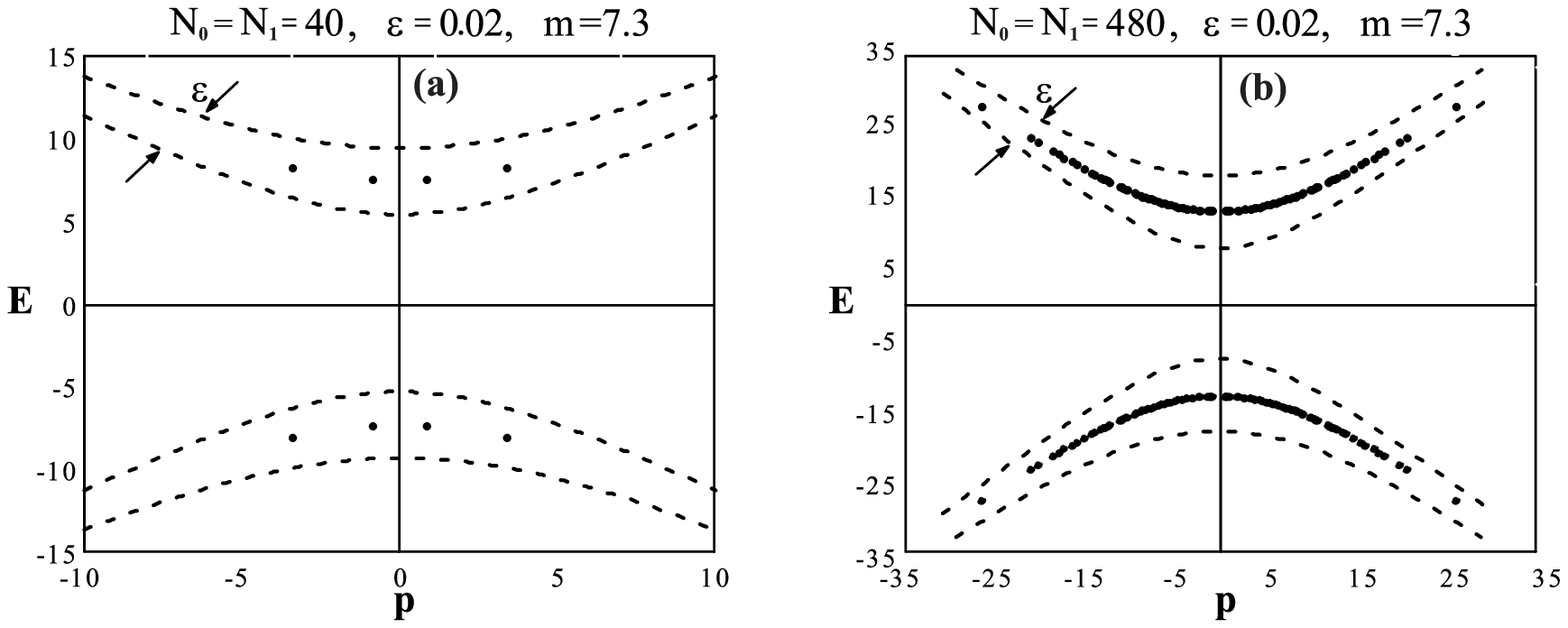}}\hfill}
\begin{center}
\begin{minipage}{14cm}
{\small 
Figure 1: The degeneracy of $m_r$ is approximated by the number of points 
$p^\mu_i$ of (\ref{masq}) located between the hyperbolas defined by
$\vert m-m_r\vert=\epsilon/2$ corresponding to the dashed lines in this 
figure for the case $m_r=7.3$. The points satisfying this condition 
are only displayed and the distance between the hyperbolas has been 
exaggerated. Since we can always find points $p^\mu_i$ close enough to 
any point laying on the lower or the upper hyperbola, both surfaces are 
completely covered when $N_\mu\to\infty$.}
\end{minipage}
\end{center}
\vskip1cm
Thus, the sum over the discrete mass--shell can be substituted 
asymptotically by a sum over all the values $p^k_i$, $k=1,2,3$, 
$i=1,2,\ldots,N_k$, and $p^0_i$ given by 
\begin{equation}
(p^0_m)^2-(p^1_j)^2-(p^2_k)^2-(p^3_l)^2=m^2, 
\label{masqpp}
\end{equation}
where $m$ is now a parameter whose numerical value is close to $m_r$.
This approximation will be used later. \\
On the other hand, a general discrete solution 
of (\ref{eqdld}) can be written as a linear combination in the degenerate 
subspace corresponding to the eigenvalue $m_r$. By using (\ref{fos}), this
combination can be written as a sum over positive values of the energy and
other sum over the negative values. 
Since the spinor space is not affected by the discretization procedure 
we can obtain the positive and negative energy bispinors following the
standard procedure. 
Thus, a plane--wave solution of (\ref{eqdld}) corresponding to the mass 
eigenvalue $m_r$ is the vector 
\begin{equation}
\Psi_{m_r}=\mathop{{\sum}^+}\limits_{p^2_q=m^2_r}
\Big[(a^1_q{\bf u}^1_q+a^2_q{\bf u}^2_q)\otimes[{\Fh}]_q\Big]+
\mathop{{\sum}^-}\limits_{p^2_q=m^2_r}\Big[
(b^1_q{\bf v}^1_q+b^2_q{\bf v}^2_q)\otimes[{\Fh}^\dagger]_q\Big],
\label{oplana}
\end{equation}
where $a^l_q$ and $b^l_q$, $l=1,2$, are complex numbers and 
\begin{eqnarray*}
&{\bf u}^1_q&=C_{mr}[1,0,p^3_l/(p^0_m+m_r),p^+_{jk}/(p^0_m+m_r)]^T,\\
&{\bf u}^2_q&=C_{mr}[0,1,p^-_{jk}/(p^0_m+m_r),-p^3_l/(p^0_m+m_r)]^T,\\
&{\bf v}^1_q&=C_{mr}[p^3_l/(p^0_m+m_r),p^+_{jk}/(p^0_m+m_r),1,0]^T, \\
&{\bf v}^2_q&=C_{mr}[p^-_{jk}/(p^0_m+m_r),-p^3_l/(p^0_m+m_r),0,1]^T.
\end{eqnarray*}
Here, $C_{mr}=(m_r/p^0_m)[(p^0_m+m_r)/2m_r]^{1/2}$ (the $u$ spinors are 
normalized to one), $p^\pm_{jk}=p^1_j\pm ip^2_k$ and $j,k,l,m$ are the 
indexes of the components of $p^2_q$ satisfying (\ref{masq}) for a 
given eigenvalue $m_r$. Eq. (\ref{masq}) is the condition that defines 
the sums given in (\ref{oplana}): positive energies on the discrete 
mass--shell correspond to the first sum and negative energies to the 
second one. Since $C_{mr}$ depends on the inverse of $p^0_m$ (eigenvalue 
of ${\Ph}^0$) the number of temporal nodes must be an even integer
to have $p^0_m\ne 0$.
\\
To obtain a relation between the temporal components of $\Psi_{m_r}$
we need to write down the explicit dependence on their spatial and 
temporal indexes maintaining the spinor structure; thus, the order 
of the tensor product of (\ref{oplana}) should be taken according 
to this. We begin by taking only $1+1$ variables to illustrate 
a standard procedure that can be generalized to more variables. The 
representation of the Dirac matrices employed 
here is $\gamma^0=\sigma_x$ and $\gamma^1=i\sigma_y$. Taking 
into account (\ref{fos}), Eq. (\ref{oplana}) can be written as
\begin{eqnarray}
(\Psi_{m_r})_q=&\Psi_{m_r}(x_j,t_k)=
\mathop{{\sum}^+}\limits_{p^2_{q'}=m^2_r}\Big[
a(p_{j'},E_{k'}){\bf u}(p_{j'},E_{k'})F^*(t_k,E_{k'})F(x_j,p_{j'})\nonumber \\
&+b(-p_{j'},-E_{k'}){\bf v}(-p_{j'},-E_{k'})F^*(t_k,-E_{k'})F(x_j,-p_{j'})
\Big], \label{oplad}
\end{eqnarray}
where the sign appearing in (\ref{fos}) has been included in the complex
constant $b_{q'}=b(-p_{j'},-E_{k'})$ and the relations between indexes are
\begin{equation}
q=j+(k-1)N_1,\qquad q'=j'+(k'-1)N_1,
\label{qqpp}
\end{equation}
where $j,j'=1,\ldots,N_1$ and $k,k'=1,\ldots,N_0$. Here, the spinors are
given by
\[
{\bf u}(p,E)={1\over{2E}}\pmatrix{m_r/\sqrt{E+p}\cr \sqrt{E+p}\cr},
\qquad
{\bf v}(-p,-E)={1\over{2E}}\pmatrix{\sqrt{E-p}\cr -m_r/\sqrt{E-p}\cr}.
\]
If we multiply (\ref{oplad}) by $F^*(x_j,p_{j\biprime})$ and sum over $x_j$ 
we obtain 
\begin{eqnarray}
\sum_{x_j}F^*(x_j,p_{j\biprime})&\Psi_{m_r}(x_j,t_k)=
\mathop{{\sum}^+}\limits_{p^2_{q'}=m^2_r}\Big[
a(p_{j'},E_{k'}){\bf u}(p_{j'},E_{k'})F^*(t_k,E_{k'})\delta_{p_{j\biprime},p_{j'}}
\nonumber \\&+
b(-p_{j'},-E_{k'}){\bf v}(-p_{j'},-E_{k'})F^*(t_k,-E_{k'})\delta_{p_{j\biprime},-p_{j'}}
\Big], \label{oplasum}
\end{eqnarray}
where we have made use of the orthogonality relations
\begin{eqnarray*}
&\sum\limits_{x_j}F^*(x_j,p_{j\biprime})F(x_j,p_{j'})=\delta_{{j\biprime}j'}=\delta_{p_{j\biprime},p_{j'}},\\
&\sum\limits_{x_j}F^*(x_j,p_{j\biprime})F(x_j,-p_{j'})=\delta_{{j\biprime},N_1-j'+1}=\delta_{p_{j\biprime},-p_{j'}}.
\end{eqnarray*}
The right-hand side of (\ref{oplasum}) is different from zero if $p_{j\biprime}$ 
is on the discrete mass--shell. By using the orthogonality relations for the spinors and
projectors we obtain that
\begin{equation}
\Psi_{m_r}(x_j,t_k)=\sum\limits_{x_{j'}}G(x_j,t_k;x_{j'},t_{k'})
\gamma^0\Psi_{m_r}(x_{j'},t_{k'}),
\label{figreen}
\end{equation}
where 
\begin{eqnarray}
G(x_j,t_k;x_{j'},t_{k'})=&\mathop{{\sum}^+}\limits_{p^2_{q\biprime}=m^2_r}
\displaystyle{{1\over{2E_{k\biprime}}}}\Big[({{\psl}_{q\biprime}+m_r})
\displaystyle{{{F^*(x_{j'},p_{j\biprime})}\over{F^*(t_{k'},E_{k\biprime})}}}
F^*(t_k,E_{k\biprime})F(x_j,p_{j\biprime}) \nonumber \\
&+({{\psl}_{q\biprime}-m_r})
\displaystyle{{{F^*(x_{j'},-p_{j\biprime})}\over{F^*(t_{k'},-E_{k\biprime})}}}
F^*(t_k,-E_{k\biprime})F(x_j,-p_{j\biprime})\Big].
\label{green}
\end{eqnarray}
Eq. (\ref{figreen}) seems to be a discrete version of Huygens' 
principle but a simple calculation shows that it is just a 
resemblance. \\
It is true that such an equation yields an identity for
$\Psi_{m_r}(x_j,t_k)$ as given by (\ref{oplad}), when $k\to k'$ 
(interchange the sums of (\ref{figreen}) and sum first 
over $x_{j'}$), but the function given by (\ref{green}) does not satisfy
$G(x_j,t_{k'};x_{j'},t_{k'})=\gamma^0\delta_{j,{j'}}$ 
and can not be considered as a discrete Green function. \\
However, according to the argument given at the beginning of this section, 
we can substitute asymptotically the sum over the discrete mass--shell 
given in (\ref{green}) by the sum over all the values of the discrete 
spatial momentum satisfying a energy--momentum relation according 
to (\ref{masqpp}), i.e, 
\begin{equation}
\mathop{{\sum}^+}\limits_{p^2_q=m^2_r}\to
\sum\limits_{\sqrt{p_j^2+m^2}=E_k}
\label{sumtosum}
\end{equation}
as $N_\mu\to\infty$ for $j=1,\ldots,N_1$, $k=1,\ldots,N_0$. 
Here, $m$ is a parameter close to $m_r$. Therefore, 
in this limit (\ref{green}) can be substituted by
\begin{eqnarray}
\tilde{G}(x_j,t_k;x_{j'},t_{k'})=&\sum\limits_{\sqrt{p_{j\biprime}^2+m^2}=E_{k\biprime}}
\displaystyle{{1\over{2E_{k\biprime}}}}\Big[({{\psl}_{q\biprime}+m_r})
\displaystyle{{{F^*(x_{j'},p_{j\biprime})}\over{F^*(t_{k'},E_{k\biprime})}}}
F^*(t_k,E_{k\biprime})F(x_j,p_{j\biprime}) \nonumber \\
&+({{\psl}_{q\biprime}-m_r})
\displaystyle{{{F^*(x_{j'},-p_{j\biprime})}\over{F^*(t_{k'},-E_{k\biprime})}}}
F^*(t_k,-E_{k\biprime})F(x_j,-p_{j\biprime})\Big].
\label{greenas}
\end{eqnarray}
It is not difficult to show that 
\[
\tilde{G}(x_j,t_{k'};x_{j'},t_{k'})=\gamma^0\delta_{j,{j'}},
\]
so that we can interpret (\ref{greenas}) as the discrete Green 
function and 
\begin{equation}
\Psi_{m_r}(x_j,t_k)=\sum\limits_{x_{j'}}\tilde{G}(x_j,t_k;x_{j'},t_{k'})
\gamma^0\Psi_{m_r}(x_{j'},t_{k'}),\qquad t_{k'}< t_k,
\label{figreenas}
\end{equation}
as the discrete version of the Huygens' principle. The condition 
$t_{k'}\le t_k$ has been added to define the retarded discrete 
propagator.
\\


\section{Propagators on the lattice} \label{propag}
Let us denote by $K(x_j,t_k;x_{j'},t_{k'})$ the $1+1$ retarded propagator. Thus
\begin{equation}
K(x_j,t_k;x_{j'},t_{k'})=\tilde{G}(x_j,t_k;x_{j'},t_{k'}), 
\qquad t_{k'}< t_k.
\label{kprop}
\end{equation}
In this section we will obtain some properties of this function. First, 
we will show the equivalence between (\ref{kprop}) and the discrete 
propagator given in \cite{Cam00a}. By using the asymptotic formula 
(\ref{fasi}) we have that 
\[
\lim_{N_\mu\to\infty} 
\displaystyle{{{F^*(0,p_{j'})}\over{F^*(0,E_{k'})}}}=
\lim_{N_\mu\to\infty}
\displaystyle{{{F^*(0,-p_{j'})}\over{F^*(0,-E_{k'})}}}=1.
\]
Therefore, 
\begin{equation}
K(x,t;0,0)=\sum\limits_{\sqrt{p^2+m^2}=E}
\displaystyle{{1\over{2E}}}\Big[({\psl}+m)
F^*(t,E)F(x,p)+({\psl}-m)F^*(t,-E)F(x,-p)\Big],
\label{kpla}
\end{equation}
where $t>0$ and the indexes have been suppressed to clarify the notation. 
Writing $F(\xi,\eta)=C(\xi,\eta)+iS(\xi,\eta)$ 
and using the parity of these functions with respect to their 
arguments, we get the components  
\begin{eqnarray*}
K_{11}(x,t;0,0)=K_{22}(x,t;0,0)=\sum\limits_{\sqrt{p^2+m^2}=E}
\Big[ -{{im}\over E}S(t,E)C(x,p)\Big], \\
K_{12}(x,t;0,0)=\sum\limits_{\sqrt{p^2+m^2}=E}
\Big[C(t,E)C(x,p)-{p\over E}S(t,E)S(x,p)\Big], \\
K_{21}(x,t;0,0)=\sum\limits_{\sqrt{p^2+m^2}=E}
\Big[C(t,E)C(x,p)+{p\over E}S(t,E)S(x,p)\Big],
\end{eqnarray*}
which are the same that those given in \cite{Cam00a}, obtained 
here under a more general scheme. As it is shown in \cite{Cam00a}, 
this discrete Green function converges to the correct continuum 
propagator and can be rewritten as the weighted sum over 
lattice paths of the checkerboard model of Feynman. \\
As usual, the condition $t_{k'}\le t_k$ in (\ref{kprop}) can be 
included as the product of the step and Green functions 
\begin{equation}
K(x_j,t_k;x_{j'},t_{k'})=\tilde{\Theta}(t_k,t_{k'})\,
\tilde{G}(x_j,t_k;x_{j'},t_{k'}), 
\label{kproptet}
\end{equation}
where $\tilde{\Theta}$ is defined by (\ref{thetap}). The Feynman 
propagator can be defined along the same lines. \\
The Hadamard product of two matrices $A$ and $B$ [defined as 
$(A\circ B)_{jk}=A_{jk}B_{jk}$] can be used to write (\ref{kproptet}) 
in matrix form. Let us define the $(2N)\times(2N)$ matrix 
\[
\tilde{\bf\Theta}=1_{2\times 2}\otimes\tilde{\Theta}\otimes O,
\]
where $N=N_0N_1$ and 
$O$ is the identity matrix for the Hadamard product 
($O_{jk}=1$, $j,k=1,2,\ldots,N_1$). Thus, taking the order for the 
tensor product used to obtain (\ref{greenas}), we have that 
\[
(\tilde{\bf\Theta})_{qq'}=1_{2\times 2}\tilde{\Theta}_{kk'}.
\] 
Now, if $\bar{\Fh}$ denotes the matrix of components
$(\bar{\Fh})_{qq'}=F(x_{j},p_{j'})/F(t_{k},E_{k'})$, the matrix 
block $\tilde{G}(x_j,t_k;x_{j'},t_{k'})$ given by 
(\ref{greenas}) takes the compact form
\begin{equation}
(\tilde{\bf G})_{qq'}=\sum\limits_{\sqrt{p_{j\biprime}^2+m^2}=E_{k\biprime}}{1\over 2}
\Big[({\Fh})_{qq\biprime}\Big({{{\psl}_{q\biprime}+m_r}\over{E_{k\biprime}}}\Big)
(\bar{\Fh}^\dagger)_{q\biprime q}+
({\Fh}^\dagger)_{qq\biprime}\Big({{{\psl}_{q\biprime}-m_r}\over{E_{k\biprime}}}\Big)
(\bar{\Fh})_{q\biprime q}\Big]
\label{gmatq}
\end{equation}
and (\ref{kproptet}) becomes the product 
$(\tilde{\bf\Theta})_{qq'}(\tilde{\bf G})_{qq'}$. This means that 
the retarded propagator is the $(2N)\times (2N)$ matrix
\begin{equation}
{\bf K}=\tilde{\bf\Theta}\circ\tilde{\bf G}.
\label{kmat}
\end{equation}
By using the relation
\[
(A\otimes B)\circ(C\otimes D)=(A\circ C)\otimes(B\circ D),
\]
it is not difficult to see that the components of ${\bf K}$ 
can be written as
\begin{eqnarray}
({\bf K})_{qq'}&=\sum\limits_{\sqrt{p_{j\biprime}^2+m^2}=E_{k\biprime}}
\displaystyle{1\over 2}
\Big[\Big((\tilde{\Theta}\circ F^\dagger_0)\otimes F_1
\Big)_{qq\biprime}\big(\displaystyle{{{\psl}_{q\biprime}+m_r}\over{E_{k\biprime}}}\Big)
(\bar{\Fh}^\dagger)_{q\biprime q}+\nonumber \\
&\Big((\tilde{\Theta}\circ F_0)\otimes F_1^\dagger
\Big)_{qq\biprime}\big(\displaystyle{{{\psl}_{q\biprime}-m_r}\over{E_{k\biprime}}}\Big)
(\bar{\Fh})_{q\biprime q}\Big]\label{kmq}.
\end{eqnarray}
Applying $(i\gamma^\mu\otimes {\Dh}_\mu-m_r{\bf 1}_{2N})$ to (\ref{kmat}) 
and using (\ref{kmq}), (\ref{derprodtet}) and (\ref{dereig}) we get  
\begin{equation}
\sum_{s}(i\gamma^\mu\otimes {\Dh}_\mu-m_r{\bf 1}_{2N})_{qs}({\bf K})_{sq'}
=i\delta_{qq'}1_{2\times 2}
\label{propin}
\end{equation}
as $N_\mu\to\infty$. Thus, the retarded propagator defined by 
(\ref{kmat}) becomes asymptotically equal to $-i$ times the inverse of 
the Dirac matrix: 
\begin{equation}
{\bf K}=-i(i\gamma^\mu\otimes {\Dh}_\mu-m_r{\bf 1}_{2N})^{-1}.
\label{kinv}
\end{equation}
Similarly, the Feynman propagator {\bf S} which propagate the 
positive frequencies toward positive times and the negative ones 
backward in time, is defined by
\begin{eqnarray}
({\bf S})_{qq'}&=\sum\limits_{\sqrt{p_{j\biprime}^2+m^2}=E_{k\biprime}}
\displaystyle{1\over{2i}}
\Big[\Big((\tilde{\Theta}\circ F^\dagger_0)\otimes F_1
\Big)_{qq\biprime}\big(\displaystyle{{{\psl}_{q\biprime}+m_r}\over{E_{k\biprime}}}\Big)
(\bar{\Fh}^\dagger)_{q\biprime q}-\nonumber \\
&\Big((\tilde{\Theta}^T\circ F_0)\otimes F_1^\dagger
\Big)_{qq\biprime}\big(\displaystyle{{{\psl}_{q\biprime}-m_r}\over{E_{k\biprime}}}\Big)
(\bar{\Fh})_{q\biprime q}\Big]\label{smq}.
\end{eqnarray}
Both {\bf S} and {\bf K} are related as in the continuum case. By 
using (\ref{thetap}) and taking into account that $\tilde{\Xi}$ and 
$\tilde{\Sigma}$ 
are skew--symmetric and symmetric matrices respectively, we have that 
$\tilde{\Theta}^T=-\tilde{\Theta}+\tilde{\Sigma}$. On the other hand, 
(\ref{sigjk}) means that 
$\tilde{\Sigma}=O$ as $N_0\to\infty$, therefore 
\[
({\bf S})_{qq'}=-i({\bf K})_{qq'}+
\sum\limits_{\sqrt{p_{j\biprime}^2+m^2}=E_{k\biprime}}{i\over 2}
\Big[
({\Fh}^\dagger)_{qq\biprime}\Big({{{\psl}_{q\biprime}-m_r}\over{E_{k\biprime}}}\Big)
(\bar{\Fh})_{q\biprime q}\Big].
\]
The second term of the right--hand side of this equation is a 
solution of $(i\gamma^\mu\otimes {\Dh}_\mu-m_r{\bf 1}_{2N}){\bf \Psi}=0$.
Since {\bf K} satisfies (\ref{propin}), {\bf S} becomes the 
inverse matrix of the Dirac operator. \\
We end this section by discussing the form that such a matrix should 
have for a finite $N_\mu$. Strictly speaking, there is no inverse 
for $(i\gamma^\mu\otimes {\Dh}_\mu-m_r{\bf 1}_{2N})$ since 
$m_r$ is an eigenvalue of $i\gamma^\mu\otimes {\Dh}_\mu$. 
However, according to the approximation procedure given at the beginning 
of section \ref{sechuy}, we can substitute $m_r$ by a value 
$m\ne m_r$ laying close enough to $m_r$ to yield a nonsingular Dirac 
operator. As a matter of fact, this value plays just the role of an 
$\epsilon$--prescription in the propagator to avoid singularities on the 
real axis.
\\
Proceeding in this way and taking the representation for the 
$\gamma$--matrices used  before, we obtain 
\begin{equation}
{\bf S}=(i\gamma^\mu\otimes {\Dh}_\mu-m{\bf 1}_{2N})^{-1}=
\pmatrix{&m{\bf\Delta}& {\bf\Delta} {\bf D}_{-}\cr &{\bf D}_{+}
{\bf\Delta} & m{\bf D}_{-}^{-1}{\bf\Delta D}_{-}\cr},
\label{sinv}
\end{equation}
where the $N\times N$ matrices ${\bf\Delta}$ and ${\bf D}_\pm$ 
are defined as follows
\begin{equation}
{\bf\Delta}=({\bf D}_{-}{\bf D}_{+}-m^2{\bf 1}_N)^{-1},\quad
{\bf D}_\pm=i({\Dh}_0\pm{\Dh}_1).
\label{kgdel}
\end{equation}
Since ${\Dh}_0$ and ${\Dh}_1$ commutes,
\[
{\bf\Delta}=(-{\Dh}_0^2+ {\Dh}_1^2-m^2{\bf 1}_N)^{-1},
\]
so that ${\bf\Delta}$ is the inverse of the discrete representation 
of the Klein--Gordon operator $-\partial_0^2+\partial_1^2-m^2$. 
It is not difficult to see that ${\bf\Delta}$, ${\Dh}_\mu$ and 
${\bf D}_\pm$ also commute. Thus, (\ref{sinv}) takes the more compact 
form 
\begin{equation}
{\bf S}=(i\gamma^\mu\otimes {\Dh}_\mu+m {\bf 1}_N)
(1_{2\times 2}\otimes{\bf\Delta}),
\label{sinvm}
\end{equation}
which is the discrete form of the well-known relationship between  
fermionic and scalar propagators. Note that this result does not 
depend on any asymptotic limit; we only need to take $N_1\ne N_0$ 
and $m\ne m_r$ to have nonsingular finite matrices.


\section{Final remarks} \label{conclus}

We end this work by pointing out the main differences between the 
discrete technique for a finite $N_\mu$ and its asymptotic 
formulation given in section \ref{sechuy}.\\
First of all, we would like to remark the similarity and 
convergence of the results of this method to the ones of 
standard field theory. As it was shown in sections \ref{sechuy} 
and \ref{propag}, the convergence of the discrete Fourier 
transform it is not enough to obtain a propagator as a sum of 
plane--wave solutions of the equations; it is needed to include 
other terms laying in the neighborhood of the mass--shell. 
This is also a requirement in the standard field theory where the
propagators are assumed off--shield and Eq. (\ref{masq}) is no 
longer satisfied. 
On the other hand, in a physical process such as dispersion of 
particles, the energy and momentum conservation are consequence
of translational invariance which is accounted by Huygens' 
principle. In our discretization scheme the translational 
invariance is lost; however later is recovered when 
the number of the lattice points goes to infinity, since the
Hermite zeros become dense in the real axis.\\
Finally, we have shown at the end of the previous section how to 
get a finite matrix form of the propagators without reference 
to any asymptotic limit. This matrices can be computed easily 
in numerical calculations of this discrete field theory.

\end{document}